\documentclass{aa}
\usepackage{graphics,natbib,amssymb}
\usepackage{rotating}
\newcommand{\rf}{\par\noindent\hangindent 15pt {}}

\begin{document}

\authorrunning{Moultaka et al.}
\titlerunning{Mapping the ISM and CSM of the GC IRS~3-IRS~13 region}
\title{VLT L-band mapping of the Galactic Center IRS~3-IRS~13 region\thanks{Resulting from ESO VLT observations of program ID numbers 71.C-0192A and 71.C-0192B}}
\subtitle{Evidence for new Wolf-Rayet type stars}
\author{J. Moultaka$^1$, A. Eckart$^1$, R. Sch\"{o}del$^1$, T. Viehmann$^1$ \& F. Najarro$^2$}
\institute{1) I Physikalishes Institut,
           Z\"ulpicher Str. 77,
           50937 K\"oln, Germany \\
           2) Instituto de Estructura de la Materia, Consejo Superior de Investigaciones Cientificas, CSIC, Serrano 121, E-28006, Madrid, Spain\\
           \email{email: moultaka@ph1.uni-koeln.de} }


\date{Received  / Accepted }

\abstract{This paper presents L-band ISAAC and NAOS/CONICA (VLT) spectroscopic observations of the IRS~3-IRS~13 Galactic Center region. The ISAAC data allowed us to build the first spectroscopic data cube of the region in the L-band domain. Using the L-band spectrum of the extinction along the line of sight towards the GC derived in a previous paper (Moultaka et al. 2004), it was also possible to correct the cube for the foreground extinction. Maps of the water ice and hydrocarbon absorption line strength were then derived. These maps are important diagnostics of the interstellar and circumstellar medium (resp. ISM and CSM): Water ices are observed in molecular clouds while hydrocarbons are usually good tracers of the diffuse ISM. These maps support our previous results that the absorption features are most probably occuring in the local Galactic center medium and can be associated with the individual sources. Moreover, turbulence seems to affect the studied region of the minispiral which appears like a mixture of a dense and diffuse medium. The comparison of the concentrations of ice and hydrocarbon absorptions around IRS~13E, IRS~6E and IRS~2 with similar concentrations at the location of the extended continuum emission around IRS~3 suggests that these sources might present outflows interacting with the surrounding ISM. It was also possible to derive Br${\alpha}$ and Pf${\gamma}$ emission line maps. The results suggest that the physical conditions of the ISM are not uniform in the observed region of the minispiral especially at the edges of the minicavity. The emission line maps allowed us to find three sources with broad lines corresponding to a FWHM deconvolved line width of about 1100 km/s and moving towards us with a radial velocity of about -300km/s. These sources are most probably new Wolf-Rayet type stars located in projection to the north and west of IRS~3. Their derived radial velocities and proper motions show that only two of them might belong to the two rotating disks of young stars reported by Genzel et al. (2003) and Levin \& Beloborodov (2003). Previously, NAOS/CONICA (NACO) data allowed us to resolve the IRS13E3 region into two components E3N and E3c (Eckart et al. 2004). The new spectroscopic NACO data show that E3c is a good candidate for beeing a Wolf-Rayet type star. In addition, three sources ($\eta$, $\zeta$ and $\gamma$) out of the eight very red sources located in the IRS13N complex also presented in Eckart et al. (2004) have been resolved spectroscopically with NACO. The spectra presented in this paper show that the red colours of the sources are probably due to extended dust emission.     

\noindent


}

\maketitle


\section{Introduction}\label{sec:intro}

The vicinity of the center of our Galaxy (at a distance of $\sim 8$kpc) makes it the ideal nuclear region that can be studied in detail. Recently, it has been shown that the central star cluster of the Milky Way is hosting a super massive black hole (SMBH) of $\sim$3.6 10$^6$ M$_\odot$ (Sch\"odel et al. 2002, Ghez et al. 2003) coinciding with the radio source Sgr A$^\star$. It is thus the best example where the environment of a central SMBH can be analysed thoroughly.\\
The central parsec is powered by a cluster of young and massive stars, the most prominent group of which is the IRS~16 complex (Blum et al. 1988, Krabbe et al. 1995, Genzel et al. 1996, Eckart et al. 1999, Cl\'enet et al. 2001) where helium emission lines have been observed in their NIR spectra. More recently, Paumard et al. (2001, 2004) have identified about 20 stars with strong He emission lines imploying integral field NIR spectroscopy. They classified them into two groups using the combination of their emission line widths and their K-magnitudes as a criterion. The intriguing result is the spatial distribution of these two groups since the narrow line stars (FWHM$\sim 225$km s$^{-1}$) are mostly found in the IRS~16 complex (diameter of about 0.15 pc) whereas the broad line stars (FWHM$\sim 1025$km s$^{-1}$) are distributed over the whole field (diameter of about 1 pc). About 20 additional early-type stars in the inner region have also been reported by Horrobin et al. (2004), Paumard et al. (2004).\\ 
The presence and especially the formation of a large number of young massive stars in the vicinity of the SMBH is problematic given the tidal field of Sgr A$^\star$ and other constraints (see discussion in Genzel et al. 2003).\\
Levin \& Beloborodov (2003) and Genzel et al. (2003) show that the early-type stars appear to be located in one or two coeval counterrotating (with respect to each other) disks at large inclination angles. The presence of such disks would then give new insights into the presence of young stars in the vicinity of the SMBH (Genzel et al. 2003, Levin \& Beloborodov, 2003). \\
In addition to these massive early-type supergiants, a population of late-type stars are identified as AGB stars (e.g. Krabbe et al. 1995, Horrobin et al. 2004). A number of embedded objects associated with dust emission are also observed in the Galactic Center stellar cluster. Examples of such objects are IRS~1, IRS~3 and also the IRS~5, 8, 10W and IRS~21 sources. These are identified as bowshock sources (Tanner et al. 2002, 2004, Rigaut et al. 2003) and are probably early-type Wolf-Rayet (WR) stars (Tanner et al. 2004) heating the dust of their environment while moving through it. \\
Recently, the IRS~13E complex has been resolved into 6 components  (Maillard et al. 2004) and the IRS~13N region, located $\sim 0.5\arcsec$ north to the center of this complex, into eight sources (Eckart et al. 2004). The latter have strongly reddened NIR colours. The authors interpret this in the framework of low luminosity bowshock sources or young stellar objects (YSO). \\

As for the interstellar medium in the central 1-2 pc region of our Galaxy, referred to as the central minispiral or the SgrA West HII region (in the following we refer to it as the minispiral) and surrounded by a torus of molecular gas called the circumnuclear ring (CNR or CND), it is shown to be composed in part of ionised gas in a low density (diffuse) medium (Lebofsky 1979, Lacy et al. 91, Paumard et al. 2004) and in part of dense molecular clouds where dust can survive (Lutz et al. 1996, de Graauw et al. 1996, Gerakines et al. 1999). Both components are responsible for the high extinction seen towards the region reaching $\sim 27$mag in the optical (Rieke et al. 1989, Chan et al. 1997, Scoville et al. 2003, Viehmann et al. 2005). On the other hand, the extinction across the central 10$\arcsec$ to 20$\arcsec$ is shown to be smoothly distributed (Scoville et al. 2003).\\
In addition, {\it intrinsic} extinction and reddening associated with individual stars are thought to occur as well in the central parsec (Blum et al. 1996, Cl\'enet et al. 2001, Moultaka et al. 2004).\\
Moreover, ISO observations have shown deep absorption features in the MIR (Lutz et al. 1996) in the direction of the Galactic Center. These are due to H$_2$O and CO$_2$ ices or aliphatic and aromatic hydrocarbons and silicates.\\
Water ice absorption at $3.0\mu$m and aliphatic hydrocarbon absorptions around $3.4\mu$m and at $3.48\mu$m have also been observed in the spectra of the bright sources located in the central parsec of the Galaxy (Jones et al. 1983, Butchart et al. 1986, McFadzean et al. 1989, Sandford et al. 1991, Wada et al. 1991, Pendleton et al. 1994, Chiar et al. 2002, Mennella et al. 2003, Moultaka et al. 2004).\\
The hydrocarbon absorption features in the MIR are caused by their CH, CC, CH$_2$ or CH$_3$ stretching modes (Duley \& Williams 1984, Butchart et al. 1986, Sandford et al. 1991, Sellgren et al. 1995, Brooke et al. 1999, Chiar et al. 2000, Grishko \& Duley 2002). The features we are interested in are the 3.4$\mu$m (a double feature at $3.38\mu$m and $3.42\mu$m) and the $3.48\mu$m absorption, both coming from stretching vibrations of the CH$_2$ and CH$_3$ groups. They are observed in the diffuse ISM and thus constitute a characteristic signature of it.\\
The water ice feature is due to vibrations in the O-H bonds. It originates from mantles of water ice that are formed by condensation of the interstellar gas onto dust particles. \\
This feature is important as it is observed in molecular clouds and is often associated with regions of star formation (e.g. Ishii et al. 1998, Brooke et al. 1999). \\

In this paper, we present L-band spectroscopic observations of the IRS~3-IRS~13 region obtained with the ISAAC and NACO instruments operating at the VLT telescopes. This allows us to obtain a detailed picture of the integrated dust distribution towards that region thanks to the first L-band data cube constructed for this goal. In the following section, we describe the observations and the data reduction. In Sect.~\ref{datacube}, we provide an analysis of the data cube constructed using the ISAAC observations and we discuss the results. The adaptive optics data obtained with NACO are presented and discussed in Sect. \ref{naco}. Finally, a summary and conclusion are provided in the last section.

\section{Observations and data reduction}\label{sec:obs}

\begin{figure}
  \resizebox{8cm}{!}{\rotatebox{0}{\includegraphics{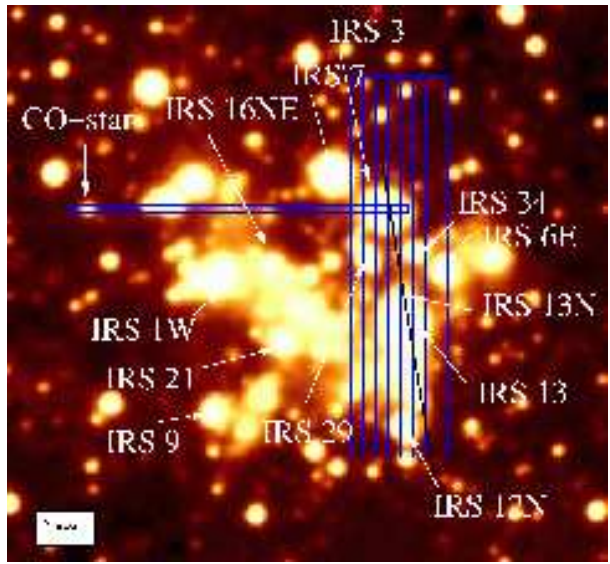}}} 
  \caption[]{ISAAC L-band image of the central parsec of the Galaxy. The slit positions as chosen for the ISAAC observations are shown in blue. The black line corresponds to the NACO slit position. The ``CO-star'' used to derive L-band spectrum of the extinction along the line of sight toward the GC lies in one of the ISAAC slit positions. }
\label{slitpositionsII.eps}
\end{figure}

L-band observations of the central $\sim$0.5 pc of the Galaxy have been undertaken with the spectrograph ISAAC located at the ESO Very Large Telescope (VLT) unit telescope UT1 (Antu), at the Paranal observatory in Chile during July 2003.
The complete set of observations with ISAAC required for our study could be obtained in a single night.\\
The observations were performed with the long-wavelength
(LWS3) and low resolution (LW) mode using the SL filter covering the wavelength range of 2.7$\mu$m - 4.2$\mu$m. The use of a 0.6$\arcsec$ slit width
implied a spectral resolution $R=\lambda/\Delta \lambda=600$ in
that wavelength domain. The optical seeing at this time was in the range between 0.4$\arcsec$ and 1$\arcsec$. To compensate for the thermal
background, separate chopped observations were carried out using chopper throws of $\sim$20$\arcsec$ along the slit of which length is of~120$\arcsec$. \\
The slit was positioned at 9 different locations shown in Fig. \ref{slitpositionsII.eps}. Eight of them map the IRS~3-IRS~13 region. The goal of such a mapping is to build a data cube of the region. For calibration purposes, a single additional slit position passes through a late-type star, labeled ``CO-star'' (see Fig. \ref{slitpositionsII.eps}) since it has been used successfully for an L-band extinction correction as described in Moultaka et al. (2004).\\
The resulting array images were divided by flat-fields, corrected for cosmic rays, for sky lines and for dispersion-related distortion. The wavelength calibration was performed using a Xenon-Argon lamp.\\ 
A chopped frame contains a positive trace image and a negative one. Two consecutive chopped frames have shifted image positions where the positive trace image of the first frame is at the same position as the negative one of the second frame. Such consecutive frames were then subtracted from each other to provide a single frame containing two negative
trace images and a positive one with twice the intensity of the
negative images. After extraction of the individual source spectra
they were corrected for wavelength dependent sensitivity, atmospheric
transmission, and telluric lines using two standard stars HR 5249
(B2IV-V) and HR 7446 (B0.5III).\\

A second night was used to observe part of the region using one slit position (see Fig. \ref{slitpositionsII.eps}) with CONICA in spectroscopic mode behind the adaptive optics system NAOS mounted on the VLT unit telescope UT4 (Yepun). The slit was positioned such that the IRS~13E3 and 3 of the IRS~13N components lay inside the slit.\\
During this night the optical seeing varied from 0.4$\arcsec$ to 0.9 $\arcsec$. The slit width used is of 86 mas implying a resolution of R=700 in the wavelength domain from $3.2\mu$m to $3.76\mu$m covered by the used L' filter. \\
For sky subtraction, the jitter technique was used consisting of taking different images with offsets between the positions. We subtracted the sky images from the object images. The resulting frames were flatfielded after a subtraction of the darks from the flatfields. They were corrected for cosmic rays, for sky lines and for dispersion-related distortion. \\
As no calibration lamp was available for wavelength calibration, the prominent atmospheric line at $3.31\mu$m and the Pf${\gamma}$ emission line at $3.739\mu$m were used instead.\\
Finally, the images were corrected for telluric lines using the spectra of the standard G0 type star HD 4306.\\

All the data reduction was performed using routines from the IRAF and MIDAS software packages.

\section{An L-band data cube of the IRS~3-IRS~13 region}\label{datacube}

\begin{figure}
  \resizebox{9cm}{!}{\rotatebox{-90}{\includegraphics{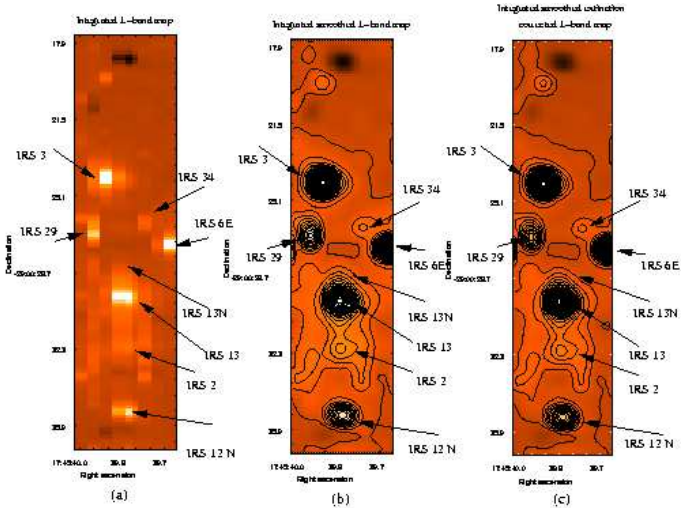}}} 
  \caption[]{{\it (a)-} The integrated L-band map of the observed IRS~3-IRS~13 region (not corrected for the foreground extinction) obtained from the ISAAC data cube. {\it (b-)} A smoothed version of the integrated map with overlaid linear intensity contours. As it is shown, one can distinguish clearly the bright sources of the region as well as the minispiral structure.{\it (c-)} A smoothed version of the integrated map corrected for the foreground extinction as explained in the text. Linear intensity contours are also overlaid for clarity which helps to notice the similarity between the two maps. They correspond to relative intensity levels since absolute values are not necessary in the context of this paper. Two consecutive contours are separated by a factor of 1/48 of the peak intensity.}
\label{cartesintegby4sm3p8IIOVI.eps}
\end{figure}

Eight of the ISAAC slit positions on the sky shown in Fig. \ref{slitpositionsII.eps} allow us to build a 3-dimensional data cube of the IRS~3-IRS~13 region (i.e. position along the right ascension. and declination as well as wavelength). This is done using the DPUSER\footnote{http://www.mpe.mpg.de/$\sim$ott/dpuser/} software. To construct such a cube, each slit image is divided by the standard star shifted in wavelength and scaled in intensity for optimum sky line suppression. The shifting allows to correct for small errors in the zeropoint of the dispersion and the scaling for possible small variations in the airmass.\\
The relative flux calibration of the slits is done 
using the L-band magnitudes published by Viehmann et al. (2005). The objects used for the relative flux density calibration of the slits are IRS~3, IRS~6E, IRS~12N, IRS~29, IRS~29N and IRS~34.\\
Due to the pixel scale of the array and to the used slit width, 
the step in declination and wavelength of the obtained cube is of 0.148$\arcsec$/pixel and of 0.6$\arcsec$/pixel in right ascension. For display purposes the images were adjusted such that the image scale is the same in both directions. The positional calibration of the cube is done using the reference coordinates of IRS~3 taken from Viehmann et al. (2005). The final calibrated image was smoothed such that the final angular resolution is of the order of 0.8$\arcsec$. Integrated L-band maps are shown in Fig. \ref{cartesintegby4sm3p8IIOVI.eps}. 
In this figure one can also distinguish the bright source IRS~3. Viehmann et al. (2005) notice an extended emission to the north-east and to the west of the compact center of this source. They interpret this feature as being an extended dust shell of a hot mass-losing star, with bow-shock-like appearance produced by the interaction with the wind from the IRS~16 cluster or SgrA$^\star$. \\

The spectrum of the L-band absorption along the line of sight towards the Galactic Center can be derived by dividing the featureless spectrum of a late-type star located well outside the minispiral area with blackbody spectrum of effective temperature T$_{eff}$= 3600K as explained in Moultaka et al. (2004). We argued in that paper that the location of this star is at a projected distance of 12.6$\arcsec$ ($\sim0.5$pc) from the center, is at the edge of the brightest part of the SgrA West HII region and does not show any excess emission in the L-band. For these reasons, it is assumed to be free of local reddening and its spectrum is then mostly but not exclusively affected by the line of sight extinction towards the Galactic Center. The spectrum of the L-band absorption is thus a good approximation of the real foreground extinction spectrum. As expected, the spectrum obtained from the present data is very similar to the one shown in Fig. 8 of the previous paper, therefore we use the same spectrum of the foreground L-band extinction. All the eight slit images are divided by this spectrum in order to construct the data cube shown in Fig.~\ref{cartesintegby4sm3p8IIOVI.eps}~c of the extinction corrected spectra.

\subsection{Optical depth maps of the ice and hydrocarbon features}\label{optdepth}

As in Sandford et al. (1991), Chiar et al. (2002) and Moultaka et al. (2004), the spectral regions around $2.8\mu$m and $3.77\mu$m are assumed to be representative of the continuum emission at these wavelengths in the L-band.\\
\begin{figure}
  \resizebox{8cm}{!}{\rotatebox{0}{\includegraphics{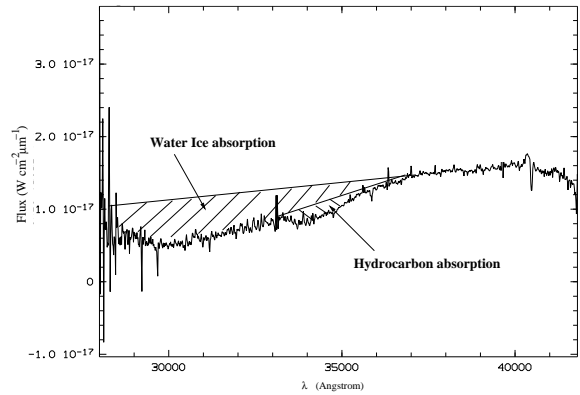}}} 
  \caption[]{L-band spectrum of IRS~29 where water ice and hydrocarbon absorptions are shown as approximated in the data cube. The approximated continua consist of straight lines from $\sim2.84\mu$m to $\sim3.77\mu$m for the H$_2$O absorption and from $\sim3.32\mu$m to $\sim3.77\mu$m in the case of the hydrocarbon feature.}
\label{IRS7cont.eps}
\end{figure}
In order to derive the optical depth maps of the water ice and the hydrocarbon absorptions from each of the data cubes described above, it is necessary to define the continua over each of the absorption features. We have approximated the continuum over the hydrocarbon absorption feature by the straight line connecting the spectrum at $\sim3.32\mu$m and $\sim3.77\mu$m; the one over the water ice absorption was approximated by a straight line connecting the spectrum at $\sim$2.84$\mu$m and $\sim3.77\mu$m (see Fig. \ref{IRS7cont.eps}). These continuum positions are good approximations as one can see by comparing such continua with the fitted ones in Moultaka et al. (2004). The water ice absorption is then approximated in the 2.8$\mu$m to 3.32$\mu$m spectral region by the area between the spectrum and the continuum over the ice feature. In the 3.32$\mu$m to 3.77$\mu$m wavelength domain, the water ice absorption is approximated by the area between the continuum over the hydrocarbon feature and the one over the water ice feature (Chiar et al. 2002) (see Fig. \ref{IRS7cont.eps}). 

Maps of the optial depths at $3.0\mu$m, $3.4\mu$m and $3.48\mu$m are also derived using the values of the continua at these wavelengths.\\

The maps are obtained using the definition of the optical depth at a given wavelength $\lambda$ by:
\begin{equation}
\tau_{\lambda}=$-ln$(\frac{F_{obs\,\lambda}}{F_{cont\,\lambda}})
\end{equation} 
where $F_{obs\,\lambda}$ and $F_{cont\,\lambda}$ are the observed and the estimated continuum fluxes at the given wavelength $\lambda$, respectively.\\
The values of the fluxes at a given wavelength are determined by taking the mean value over 60 pixels ($\sim 85.8$nm) around $\sim2.84\mu$m, 10 pixels ($\sim 14.3$nm) around $\sim3.32\mu$m and $\sim3.77\mu$m and over 40 pixels ($\sim 57.2$nm) around $3.0\mu$m, $3.4\mu$m and $3.48\mu$m.

When deriving the optical depth values over the IRS~3-IRS~13 region, some conditions have been imposed to avoid artifacts of the division procedure and 
in regions of low flux densities. Thus, the optical depth value is set to zero if the signal is less than 3$\sigma$ of the sky background noise in the data.\\

Since the resulting observed and extinction corrected maps are very similar, we show only the extinction corrected ones in Fig.~\ref{carteby4p8IIOVIIVExtCorr.eps}. In this figure are shown the integrated L-band map of the IRS~3-IRS~13 region (Fig.~\ref{carteby4p8IIOVIIVExtCorr.eps}~a), the optical depth integrated maps over the ice feature from $2.84\mu$m to $3.77\mu$m (Fig.~\ref{carteby4p8IIOVIIVExtCorr.eps}~b) and over the hydrocarbon feature from $3.32\mu$m to $3.77\mu$m (in Fig.~\ref{carteby4p8IIOVIIVExtCorr.eps}~c). The maps obtained after deriving the optical depth values at the wavelengths of $3.0\mu$m, $3.4\mu$m and $3.48\mu$m are shown in (d), (e), (f) of the figure. 

The striking first result, as reported previously, is that the non corrected maps and their extinction corrected homologues are very similar. This shows that the absorption features probably take place in the local Galactic center medium and that, when they are present at the location of a bright source, they can be associated with the source. This result supports our previous finding concerning the absorption features observed in the direction of the bright sources of the central half parsec (Moultaka et al. 2004).\\
Moreover, the maps of the integrated ice absorption feature (maps b) and of the absorption at 3.0$\mu$m (maps d) are very similar and those of the integrated hydrocarbon feature (maps c), the 3.4$\mu$m (maps e) and the 3.48$\mu$m (maps f) are similar as well. This shows that the values of the optical depths obtained at the single wavelengths are well representative of the shape of the absorption features over the whole absorption band.\\
All the previous results also show that our continuum derivation is reliable. 

\begin{figure}
  \resizebox{8cm}{!}{\rotatebox{0}{\includegraphics{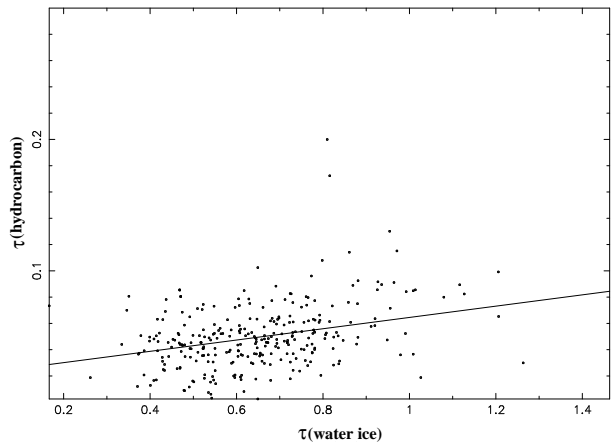}}} 
 \caption[]{Optical depth of the hydrocarbon versus the ice absorption features. This figure plots the intensities of all pixels of the water ice and hydrocarbon optical depth maps, at which the S/N ratio in the integrated L-band map, is higher than 10.} 
\label{diagramTauicehydmap.eps}
\end{figure}

The comparison of the ice (b and d) and hydrocarbon (c, e and f) optical depth maps show for the first time that the two features are similarly distributed in the region and are prominent in the studied minispiral area. Indeed, the peaks and minima in both absorptions are located at the same positions in the different maps. A vague correlation between the optical depth values of the two absorption features observed towards the bright sources of the central half parsec was found in our previous paper Moultaka et al. (2004), (see Fig. 19 of that paper). This correlation shows up as well in the present data and in the studied region. In Fig.~\ref{diagramTauicehydmap.eps}, are plotted the optical depth values of the two absorption features at different pixels of the maps where the signal to noise ratio is higher than 10, implying small errors on the optical depth measurments. These pixels include the bright sources of the region and a large part (the central part) of the observed minispiral area. The correlation coefficient obtained without considering the uncertainties is 0.36 which is in agreement with the previous value of 0.43 where the error bars were taken into account (Moultaka et al. 2004). This correlation suggests that the ISM presents itself as a mixture (possibly clumpy as indicated by the maps) between a dense dusty and a diffuse ionised media. In addition, the turbulent nature of the ISM (Paumard et al. 2004) may also be of importance.

The northern part of IRS13, where the presence of very red sources has been reported by Eckart et al. (2004), shows higher water ice and hydrocarbon absorptions than the center of the IRS~13 complex. This is in agreement with these sources being more embedded than those at the center of the IRS~13 complex.

On the other hand, one can notice in the optical depth maps a peak in the ice absorption feature at the north-eastern and the western sides of IRS~3. These peaks possess homologues in the hydrocarbon absorption feature maps. In addition, the position of IRS~3 shows a deficiency in the H$_2$O absorption which suggests that the ISM has been swept up by the stellar wind at this position. The peaks can be interpreted as being the trace of the circumstellar medium at the edga of the extended dust shells of this star. This also suggests that the west-side bowshock-like extended emission which shows up in the integrated L-band map (Fig.~\ref{cartesintegby4sm3p8IIOVI.eps}) is real and can be created by the mass-losing star during its interaction with the surrounding medium.\\
A similar situation occurs north and south of IRS~13 and IRS~6E and north-west and south-east to IRS~2 where peaks in the ice and hydrocarbon absorptions show up; moreover, the absorption maps present also a deficiency at the location of IRS~13, IRS~6E and IRS~2 as it is the case for IRS~3. This is an intriguing result which suggests that these sources could also be mass-losing stars interacting with their environment and presenting an outflow-like shape.\\
Moreover, these results indicate a close spatial correlation between these sources, especially IRS~13, and the surrounding material of the minispiral.
    
Finally, both IRS~34 and IRS~29 locations show high water ice and hydrocarbon absorption peaks that can be associated with these sources.

\subsection{Hydrogen emission line maps}\label{emlines}

The maps of the Pf${\gamma}$ and Br${\alpha}$ emission lines are constructed by deriving the relative flux density integrated over each line (from $3.722\mu$m to $3.744\mu$m for Pf${\gamma}$ and from $4.028\mu$m to $4.061\mu$m for Br${\alpha}$). The continuum is taken as the average of the continuum on the red and the blue sides of the line taken over 16 pixels ($\sim 22.8$nm ) for Pf${\gamma}$ and 24 pixels ($\sim 34.3$nm) for Br${\alpha}$ on each side.\\
Since we compute the relative fluxes (i.e. $\frac{F_{line}-F_{cont}}{F_{cont}}$) and in order to avoid artifacts in the maps (see Sect. \ref{optdepth} above), we impose similar conditions as the ones used for the optical depths of the water ice and hydrocarbon features.

The Pf${\gamma}$ and Br${\alpha}$ emission line maps are shown in Fig. \ref{carteby4p8IIOVIIVExtCorr.eps} (g) and (h) where the spectra have been corrected for the foreground extinction. These maps are similar to those not corrected for the line of sight extinction as one could expect because the wavelength region of the emission lines in the spectra is not heavily affected by extinction.
 
The maps of the two emission lines trace well the minispiral and are very similar in that area. They show high line strengths in the south-eastern part which can be explained probably by the high density of this region being located just at the south-western edge of the minicavity and possibly associated with the late-type star IRS~12N (Genzel et al. 2000).
In Fig.~\ref{carteby4p8IIOVIIVExtCorr.eps}~(i), we show the map of the Br${\alpha}$ to Pf${\gamma}$ ratio. This map does not show a constant ratio over the minispiral region; the line ratio varies from 0.12 to 1.75 and shows peaks in the south-western part associated with the south-western edge of the minicavity. Peaks of the ratio show up as well in the region located to the north-west of the minicavity with values of about 1.8 and at the location of the IRS~2 and IRS~13 complex regions with values of 1.6 and 1.4, respectively. The IR colour-temperature map of the minispiral obtained by Cotera et al. (1999) shows peaks of the dust temperature at the locations of IRS~13 and IRS~2 corresponding to values of about 220K. As shown by Hummer \& Storey (1987) for temperatures $\geq$ 1000 K, the Br$\alpha$/Pf$\gamma$ ratio can be equal in media of high temperature and low density and vice versa. Our Br$\alpha$/Pf$\gamma$ ratio maps are in agreement with the finding of Cotera et al. (1999) if the density at the locations of IRS~13 and IRS~2 is lower than in the neighbouring regions. The peaks of the line ratio at the south- and north-western sides of the minicavity can then be explained by lower temperatures and higher densities well justified by their location at the edges of the minicavity and possibly associated with IRS~12N and IRS~29. \\

\subsubsection{New WR-type stars in the Galactic center}
\begin{figure}
  \resizebox{6cm}{!}{\rotatebox{0}{\includegraphics{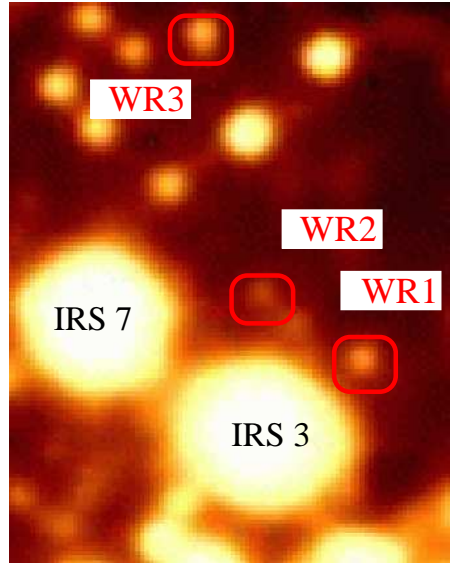}}} 
  \caption[]{L-band ISAAC image of the IRS~7-IRS~3 region with the positions of the three WR-type stars WR1, WR2 and WR3 where WR1 is the IRS~7W source identified by Ott (2004) and Paumard et al. (2001,2004).}
\label{WRpositions.eps}
\end{figure}

\begin{figure}
  \resizebox{8cm}{!}{\rotatebox{0}{\includegraphics{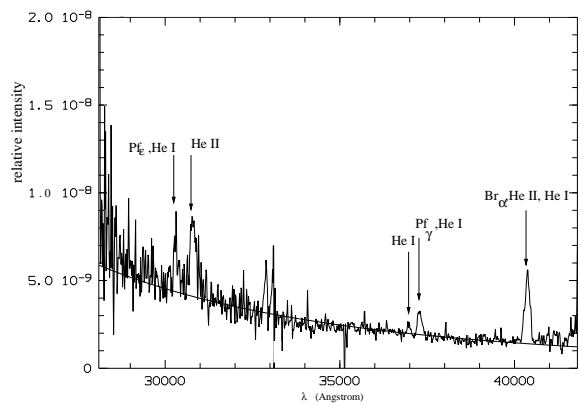}}}  
  \caption[]{L-band spectrum corrected for the line of sight extinction of the Wolf-Rayet type star WR1. The broad Helium emission lines are also shown. The shape of this spectrum agrees well with a blackbody temperature $\geq$ 30000K. }
\label{WR1FRVN11ExtCorr.eps}
\end{figure}

\begin{figure}
  \resizebox{8cm}{!}{\rotatebox{0}{\includegraphics{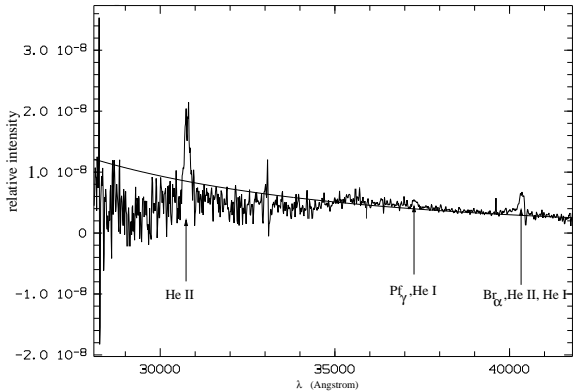}}}  
  \caption[]{Extinction corrected L-band spectrum of the Wolf-Rayet type star WR2 where Helium emission lines are distinguished. A blackbody spectrum with effective temperature of 30000K is also shown. The shape of the spectrum can be well fitted with this continuum except in the blue part where the excess of absorption is probably due to an underestimation of the extinction towards this position. }
\label{WR2sumFRIIN7ExtCorr.eps}
\end{figure}

\begin{figure}
  \resizebox{8cm}{!}{\rotatebox{0}{\includegraphics{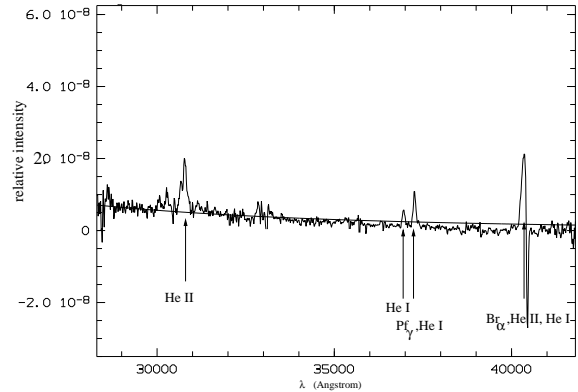}}}  
  \caption[]{Observed L-band spectrum (not corrected for the foreground extinction) of the Wolf-Rayet type star WR3 where helium emission lines are distinguished. It is well-fitted by the spectrum of a blackbody of temperature $\geq$30000K. The correction for the extinction along the line of sight as derived in our previous paper Moultaka et al. (2004) seems to overestimate the real extinction towards the position of this object.}
\label{WR3sumFRIN8.eps}
\end{figure}

The analysis of our emission line maps reveals three Wolf-Rayet stars two of which are detected for the first time. Indeed, in the northern part of the emission line maps, three sources of high emission show up, but are very faint in the integrated L-band map. They are located at the northern and the western sides of IRS~3 and correspond to three faint sources that can be distinguished in the continuum images of Fig. \ref{WRpositions.eps} (red circles labeled WR1, WR2 and WR3). Their emission lines are much broader than those of the minispiral area which have typical FWHM of 200 km/s resulting, when unresolved, from multiple individual velocity gas components which characterize the gas and have line widths of about 50 km/s (Paumard et al. 2001, 2004). The L-band spectra are shown in Figs. \ref{WR1FRVN11ExtCorr.eps}, \ref{WR2sumFRIIN7ExtCorr.eps} and \ref{WR3sumFRIN8.eps}. 
In these figures, one can distinguish several broad emission lines. All of them are intriguingly blueshifted by about -300 $\pm$ 100 km/s. Because of the broad width of the lines (the FWHM corresponds to a velocity of about 1100 km/s) and the low spectral resolution, no better precision on the radial velocities can be provided.
Taking into account the blue shift, the new emission lines correspond to HeI, HeII and H/HeI,HeII lines as labeled in their spectra. If we consider
the strong line at 
3.09$\mu$m as a Pf$\delta$ line, it would be in disagreement with the position of the other 
lines and would imply a higher blueshift of these objects which is unlikely. 
This line is also very prominent in the spectrum of WR1 located west of IRS~3, which is a HeI emission line star already identified by Najarro et al. (1997), Ott (2004) and Paumard et al. (2001,2004) and called IRS~7W. If the 3.09$\mu$m line would be due to CIII or CIV lines, then very strong CIII and CIV lines would also be expected to show up in the K-band spectrum of this star which is not the case (see Najarro et al. 1997 and Ott 2004).
Consequently, at this wavelength, only a HeII (6-7) line would fit. \\
On the other hand, the strong similarity between the spectra of WR1 and WR147 shown in Morris et al. (2000) and classified as a WN8 star, points towards a slightly lower excitation of WR1 which could prevent the HeII line at 2.189$\mu$m (fairly weak in Morris et al. 2000) from appearing in emission, while a strong HeII emission line at 3.09$\mu$m can still be observed. 
Using a radiative transfer model, Najarro et al. (1997) show indeed that HeII emission lines can be expected in the spectrum of WR1. \\
Moreover, the K-band spectrum of WR1 shows a highly depleted hydrogen (Najarro et al. 1997).
Consequently, we can conclude that the hydrogen lines {in the present L-band spectra} are also contaminated by HeI and HeII or are pure HeI neighbouring emission lines that cannot be separated at the present spectral resolution. Indeed, for higher transitions than the Bracket ones, the HeI atoms are very hydrogenic.\\ 
Najarro et al. (1997) also find a radial velocity of -250 km/s in agreement with our value and with the value obtained by Ott (2004).\\ 
The extinction corrected spectra (in the case of WR1 and WR2) are fitted by a blackbody spectrum with T$_{eff}\geq 30000$ K. Since these temperatures and the present wavelength domain correspond to the Rayleigh-Jeans regime, the temperature of the best fitted blackbody spectrum will be taken as a lower limit. A residual absorption in the blue part of the WR2 extinction corrected spectrum shows up. This is probably due to an excess of the extinction towards this position. The spectrum of WR3 shown in Fig. \ref{WR3sumFRIN8.eps} is not corrected for the foreground extinction because the corrected spectrum is very blue. This is most probably due to a lower extinction toward this northern location. Indeed, the NIR colours of this object listed in table \ref{WRtab} and obtained from ISAAC (for the L- and M-bands) and NACO (for the H- and K-bands) images as described in Viehmann et al. (2005) show that it is bluer than the others.\\
In addition, the K-band magnitudes of the 3 stars are similar to those of the broad emission line stars observed by Paumard et al. (2001,2004).\\  
Considering all these results, we conclude that these sources are most likely WR-type stars with strong stellar winds and probably high mass loss rate. As the L-band spectra of WR2 and WR3 are very similar to that of the early-type star WR1, this supports our conclusion.\\
Furthermore, the similarity of the spectrum of WR1 with the one of the WN8 star WR147 of Morris et al. (2000) shows that this star seems to be consistent with a WN8/9 spectral type although one can not rule out that it is on its way from a WN8/9 to a WC9/10 spectral type.\\

Using the K-band NACO images obtained from four epochs of NACO observations between May 2002 and July 2004 (Sch\"odel et al. in preparation), we derived the proper motions of the three WR stars (see Table \ref{WRtab2}). In this table are also listed the measured radial velocities, the angle between the normal to the disk and the velocity vector of a given star, the maximal radial velocities which allow these stars to be bound to the SMBH assuming that they belong to one of the disks. Note that if the radial veloctities of the stars are higher than the maximal radial velocities, this would not imply that they are not bound to the central stellar cluster. The velocities of the WR1 and WR2 stars fit well, within the uncertainties, the counterrotating disks of young stars determined by Genzel et al. 2003 and Levin \& Beloborodov 2003; where WR1 belongs to the clockwise rotating disk and WR2 to the counter-clockwise disk. However, WR3 seems not to belong to the counter rotating disk even though it is moving counter clockwise, but it still may, given the uncertainties on its radial velocity. 
More accurate radial velocities measurements are necessary to confirm our results.    

\begin{table*}[htbp]
\small
\begin{center}
\begin{tabular}{rccccccc}
\hline
Source & H  & K   & L & M & H-K & K-L & L-M \\
\hline
WR1  & 14.25 $\pm$ 0.25 & 11.73 $\pm$ 0.25 & 10.08$\pm$ 0.15  & 9.37$\pm$ 0.15  & 2.52 $\pm$ 0.5& 1.65 $\pm$ 0.4& 0.72$\pm$ 0.3 \\
WR2  & 15.24$\pm$ 0.25  & 12.90$\pm$ 0.25  & 11.31$\pm$ 0.15  & 10.02$\pm$ 0.15  & 2.34 $\pm$ 0.5& 1.58$\pm$ 0.4 & 1.29$\pm$ 0.3 \\
WR3  & 13.43$\pm$ 0.25 & 11.64 $\pm$ 0.25 & 9.84$\pm$ 0.15  & 9.22$\pm$ 0.15  & 1.79 $\pm$ 0.5& 1.80$\pm$ 0.4 & 0.61$\pm$ 0.3 \\
\hline
\end{tabular}
\end{center}
\caption{Magnitudes and colours of the three newly found Wolf-Rayet stars. 
}
\label{WRtab}
\end{table*}

\begin{table*}[htbp]
\small
\begin{center}
\begin{tabular}{rccccc}
\hline
Source & $\mu_\alpha$ (km/s)& $\mu_\delta$ (km/s)  & $v_r$ & $v_{r\,max}$ & $\theta$ (degrees)\\
\hline
WR1  &  -61$\pm$24 & -127$\pm$61 & 260$\pm$100 & 296$\pm$49& 85$\pm$11\\
WR2  &  135$\pm$32 & 41$\pm$47 & 310$\pm$100 & 322$\pm$35 & $ 115\pm$14\\
WR3  &  -129$\pm$39 & -4$\pm$53 & 332$\pm$105 & 174$\pm$44 & 74$\pm$12\\
\hline
\end{tabular}
\end{center}
\caption{Proper motions, radial velocities, orbital inclinations ($\theta$) to the normal to the disks of the newly found Wolf-Rayet stars and maximal line-of-sight velocities allowed for the stars to still be bound to the SMBH under the assumption that they are part of one of the disks.}
\label{WRtab2}
\end{table*}

\section{Resolving the IRS13-IRS13N complex with the adaptive optics NACO system}\label{naco}
\begin{figure*}
 \resizebox{16cm}{!}{\rotatebox{0}{\includegraphics{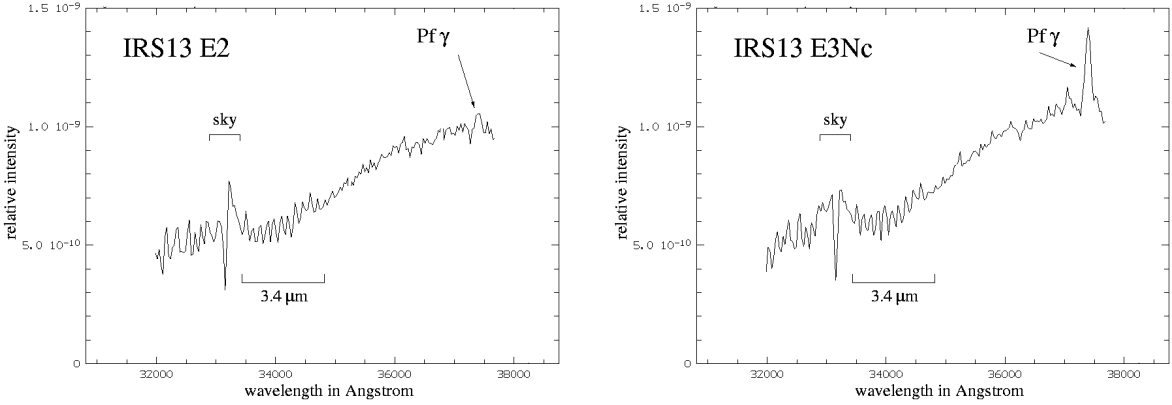}}} 
  \caption{{\it L'}-band NACO spectra of IRS~13~E2 ({\it left}) and IRS~13~E3Nc ({\it right}) where both E3N and E3c contribute to the emission. The ``sky'' and ``3.4$\mu$m'' labelled regions correspond to the partially corrected telluric methane absorption.}
\label{fig3}
\end{figure*}

In a previous paper (Eckart et al. 2004), H-, K$_s$- and L-band diffraction limited images of the Galactic Center have been obtained with the adaptive optics NAOS/CONICA camera. As shown in that paper, the IRS~13 complex has been resolved into 4 components (see Fig. 2 of that paper). In addition to the well known components E1 and E2 (Paumard et al. 2001, Maillard et al. 2004), the E3 component appears as a double structure in the K$_s$-band image. The components of this structure have been labeled E3N (for North) and E3c (for center). The E3c component is very red (K-L=4.24 and H-K=4.05) and is also the faintest among the four components in the H- and K-bands and the brightest in the L-band. In Maillard et al. (2004), the IRS~13E complex has been resolved into 6 components. The IRS~13~E4 component is the brightest in H and K after IRS~13E1 and E2. As the IRS~13~E3N of Eckart et al. (2004) has similar magnitudes, it corresponds most probably to the E4 component of Maillard et al. (2004). Moreover, the IRS~13~E3 component of Maillard et al. (2004) has been resolved into 2 components called E3A and E3B, the position of these sources is coinciding with our E3c component. Using the results of the fit of SED fitting, the authors argue that both E3A and E3B are probably WR-type stars.\\
The present NACO spectrum of the E3 region (E3N + E3c) (Fig. \ref{fig3}) shows a prominent Pf$\gamma$/HeII emission line that is very faint in the IRS~13E2 component as shown in the figure. This supports the result that the E3 component is the broad He emission-line star identified in the IRS~13 region by Paumard et al. (2001) and previously by Krabbe et al. 1995 and Blum et al. (1996). As the E3c source is the faintest in the K-band with a K-magnitude fainter than 10.6 mag, we argue that this resolved component might be a WR-type star (see discussion in Paumard et al. 2001 and the introduction of the present paper) as was also suggested in the former paper by Eckart et al. (2004). This would be in agreement with the result of Maillard et al. (2004). 

The previous NACO images have also revealed a new complex of IR red colour excess sources located at about 0.5'' north of the IRS~13 cluster. Eight sources have been resolved in the deconvolved images and labeled $\alpha$ through $\eta$ as shown in Figs. 1 and 2 of Eckart et al. (2004).
\begin{figure}
 \resizebox{8cm}{!}{\rotatebox{0}{\includegraphics{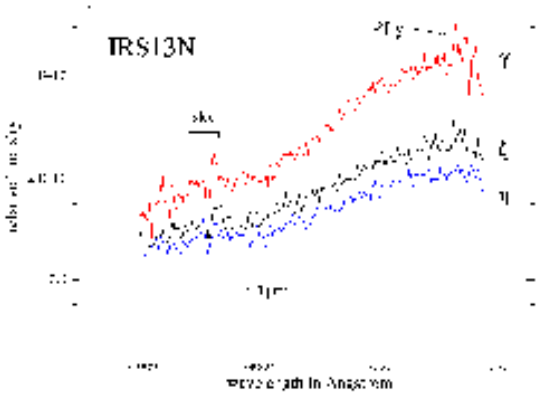}}} 
  \caption{NACO {\it L'}-band AO spectra of the IRS~13N sources $\eta$, $\zeta$ and $\gamma$ obtained with NACO instrument. The ``sky'' and ``3.4$\mu$m'' labelled regions correspond to the partially corrected telluric methane absorption.}
\label{fig2}
\end{figure}

In order to explain the unusual colours of these sources, the authors provide three different interpretations: 1- the sources could be bright young stars that are deeply embedded in the gas and dust of the minispiral or simply located behind a dense clump of gas and dust. 2- The second possibility is that they could be O-or B-type stars heating the gas and dust in their vicinity. In this case, they could be low luminosity analogues of the bowshock sources and in both cases, extinctions well above A$_V\sim$ 30 $mag$ are needed. 3- The third interpretation is that they may be young stellar objects as they show similar luminosities ($\sim 10^3$L$_\odot$) in addition to their red colours ($K-L\sim 4$ to $5$ mag). Moreover, when corrected for the visual extinction value A$_V\sim$30 mag toward the Galactic Center, their colours are shifted to the region of YSO and Herbig Ae/Be stars in the Colour-Colour diagram (Ishii et al. 1998) (see Fig. 3 of Eckart et al. 2004). In addition, in Moultaka et al. (2004), the ISAAC spectrum of the IRS~13N region has been well fitted with a reddened blackbody continuum of temperature T=1000K and the corrected spectrum for the foreground extinction is redder than the one of the IRS~13 cluster especially longward of 3.5$\mu$m. This shows that the L-band excess is due to the emission of warm (T$\sim$1000K) dust. \\
The NACO observations have resolved three of the very red new sources of the IRS13N region. These are the sources $\eta$, $\zeta$ and $\gamma$ laying along the same slit. The spectra show a redder continuum from $\eta$ to $\gamma$ as shown in Fig. \ref{fig2} which agrees well with the colours of these sources (Eckart et al. 2004). This result indicates that the colours of these sources are affected by dust emission. Moreover, the lack of Pf$\gamma$ line emission in these spectra which is prominent in the integrated lower spatial resolution ISAAC spectrum of the overall IRS~13N region as shown in Moultaka et al. (2004) suggests also that this line emission is likely due to an extended nebular emission. Alternatively, it may originate from sources $\epsilon$ or $\delta$.
  
\section{Conclusion and discussion}

The presented L-band ISAAC observations allowed us to build the first spectroscopic L-band data cube of the IRS~3-IRS~13 region. Using the spectrum of the foreground extinction towards the Galactic Center derived in a previous paper Moultaka et al. (2004), we constructed the extinction corrected data cube of the same region.\\
Maps of water ice absorption at $3.0\mu$m and of the hydrocarbon absorptions at $3.4\mu$m and $3.48\mu$m were then derived as well as the Pf$\gamma$ and Br$\alpha$ emission line maps. The three absorption features - the water ice and the hydrocarbons - are good tracers of the ISM, molecular clouds and the diffuse medium, respectively.\\
The analysis of the absorption feature maps showed that the features are present almost over the entire region and are surprisingly well correlated. This suggests that the ISM in the studied area could be affected by turbulence or could be a mixture of a dense dusty and diffuse ionised medium. On the other hand, the observed maps and the extinction corrected ones are very similar and the absorption features show varying strengths all over the region, just as well in the observed maps as in the extinction corrected maps. This consolidates our previous finding that these absorptions are most probably taking place locally at the position of the Galactic Center and can be associated with the bright sources when they are found at their location.\\
The extended emission north-east and west to the probably hot, young, mass-loosing star IRS~3 are well traced in the ice and hydrocarbon maps. Since similar absorption distributions are also observed in the maps around the IRS~6E and the IRS~12N sources, a tempting question would be whether these sources are also mass-losing stars where outflows are interacting with their environment. The continuum images obtained from Adaptive Optics observations of the region (e.g. Genzel et al. 2003) show that IRS~6E and IRS~12N are embedded in highly extended emission and can hardly be distinguished from the minispiral. Moreover, the position of the IRS~13 complex shows similar concentrations of the absorptions north and south to IRS~13E were we have just located the position of a WR-type star. This also supports the idea that such concentrations may be created by the strong winds of neighbouring massive stars.\\
These results also stress the idea that a high spatial correlation occurs between these sources and the minispiral material.\\ 
The ISAAC emission line maps have revealed three WR-type stars located to the north and west of IRS~3. One of these has already been identified as being a Wolf-Rayet type star from its K-band spectrum and the two remaing stars are reported in the present paper for the first time. The L-band spectra of the three sources show broad blueshifted helium emission lines. They correspond to FWHM line widths of about 1100 km/s and radial velocities of about -300 km/s. A better accuracy of these quantities should be obtained with spectroscopy in the NIR domain, where other WR-type stars have been identified.\\ 
Two of the new WR stars, WR1 and WR2, could belong to the two couter-rotating disks of young stars, as shown by their proper motions and radial velocities within the uncertainties. The northern star WR3 does not seem to be part of any of the two disks; but NIR spectroscopy should be undertaken to confirm these conclusions.\\
The NACO L-band observations allowed us to resolve three of the IRS~13N compact sources and the IRS~13E2 and E3Nc components of the IRS~13E complex. The absence of prominent emission lines in the IRS~13N sources, $\eta$, $\zeta$ and $\gamma$, shows that the red colours of these objects are probably due to dust. It also shows that the recombination line emission is mostly either due to extended nebular emission or due to the remaining IRS~13N sources $\epsilon$ or $\delta$.\\
On the other hand, the comparison of the NACO spectra of IRS~13E3 (IRS~13E3N and IRS~13E3c) and of IRS~13E2 suggests that the previously resolved E3c component can probably be identified with a dusty WR star. \\
This brings the number of new WR-type stars reported in this paper to three which increases the number of massive hot WR stars in the central parsec by about 17\% over the total number obtained by Paumard et al. (2001,2003) and Tanner et al. (2004) including the five bowshock sources of the northern arm. It supports the conclusion of Tanner et al. (2004) that these sources are underestimated in the central parsec of the Milky-Way.

\begin{figure*}
\begin{center}

  \resizebox{6.7cm}{!}{\rotatebox{90}{\includegraphics{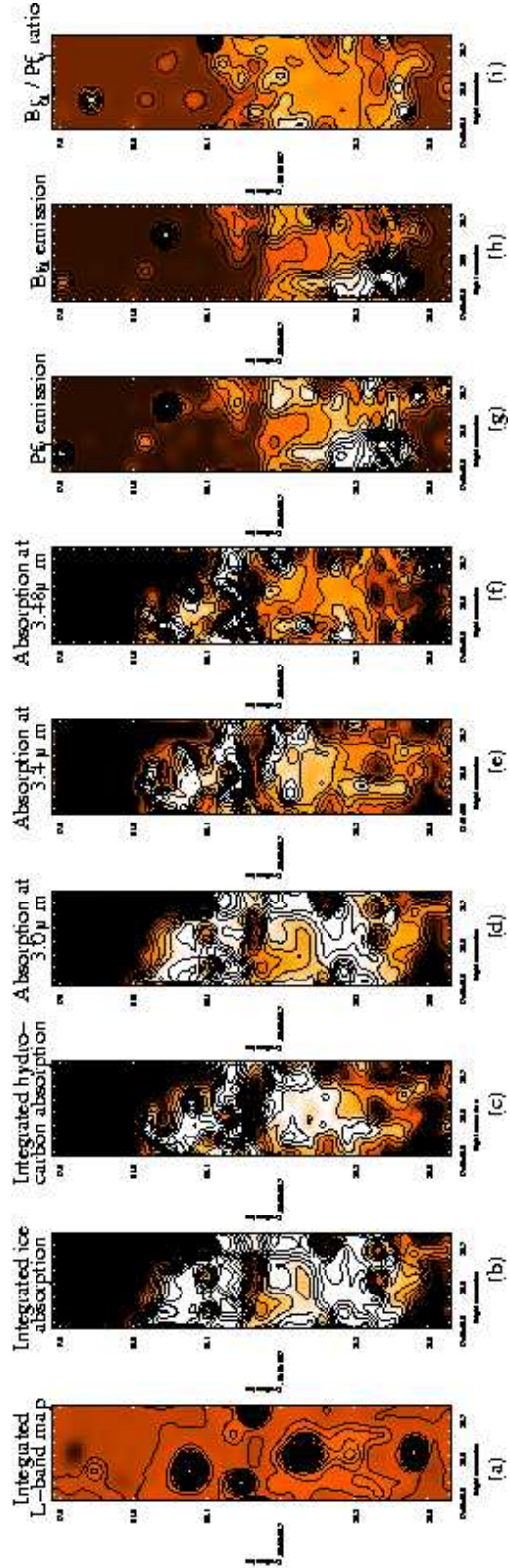}}}  
  \caption[]{\small{Maps of the IRS~3-IRS~13 GC region corrected for
  extinction along the line of sight. All over the maps are also overlaid contours of the same images for clarity. {\it (a)-} Smoothed integrated
  map over the L-band wavelength domain. The contours correspond to relative intensity levels. {\it (b)-} Smoothed version
  of the optical depth map of the integrated water ice absorption over
  the 2.84$\mu$m to 3.77$\mu$m spectral domain as explained in the
  text. Contour values: 0.07, 0.13, 0.21, 0.27, 0.34, 0.41, 0.48, 0.55, 0.62, 0.69, 0.75, 0.82, 0.89, 0.96. {\it (c)-} Smoothed version of the optical depth map of the
  integrated hydrocarbon absorption feature from 3.32$\mu$m to
  3.77$\mu$m. Contour values: 0.020, 0.028, 0.036, 0.044, 0.052, 0.060, 0.068, 0.075, 0.083, 0.091, 0.099, 0.107, 0.115 {\it (d)-} Smoothed map of the optical depth
  distribution at 3.0$\mu$m. Contour values: 0.17, 0.34, 0.51, 0.68, 0.85, 1.02, 1.19, 1.36, 1.53, 1.70, 1.87, 2.04, 2.21, 2.38 {\it (e)-} Smoothed map of the optical
  depth distribution at 3.4$\mu$m. Contour values: 0.04, 0.05, 0.07, 0.08, 0.09, 0.11, 0.12, 0.13, 0.15, 0.16, 0.18, 0.19, 0.20 {\it (f)-} Smoothed map of the
  optical depth distribution at 3.48$\mu$m.Contour values: 0.010, 0.02, 0.026, 0.031, 0.037, 0.043, 0.048, 0.054, 0.060, 0.066, 0.072, 0.077, 0.083 {\it (g)-} Smoothed version
  of the Pf$\gamma$ emission line map. Contour values: 0.02, 0.04, 0.06, 0.08, 0.11, 0.13, 0.15, 0.17, 0.19, 0.21, 0.24, 0.26, 0.28, 0.30, 0.32 {\it (h)-}  Smoothed version
  of the Br$\alpha$ emission line map. Contour values: 0.044, 0.11, 0.19, 0.25, 0.33, 0.40, 0.47, 0.54, 0.61, 0.68, 0.76, 0.83, 0.90, 0.97, 1.04, 1.11, 1.18 {\it (i)-} Smoothed version
  of the Br$\alpha$/Pf$\gamma$ ratio. Contour values: 0.12, 0.45, 0.77, 1.10, 1.42, 1.75, 2.07, 2.40, 2.72, 3.05, 3.38, 3.70, 4.03.}}
\label{carteby4p8IIOVIIVExtCorr.eps}
\end{center}
\end{figure*}

\begin{acknowledgement}
This work was supported in part by the Deutsche Forschungsgemeinschaft 
(DFG) via grant SFB 494.\\
F.N. acknowledges PNAYA2003-02785-E and AYA2004-08271-C02-02 grants and
 the Ramon y Cajal program.

\end{acknowledgement}
\appendix


\bibliographystyle{apj}

%
%
%
%
%
%
 
\rf{Blum, R. D., Sellgren, K., Depoy, D. L. 1988, AJ 112}
 
%
%
\rf{Blum, R. D., Sellgren, K., Depoy, D. L. 1996, ApJ 470, 864 }
%
%
%
\rf{Brooke, T. Y., Sellgren, K., Geballe, T. R. 1999, ApJ 517, 883 }
%
%
%
\rf{Butchart, I., McFadzean, A. D., Whittet, D. C. B., Geballe, T. R., Greenberg, J. M. 1986, A\&A 154 }
\rf{Chan, Kin-Wing, Moseley, S. H., Casey, S., Harrington, J. P., 
    Dwek, E., Loewenstein, R., Varosi, F., Glaccum, W. 1997,
    ApJ 483, 798 }
\rf{Chiar, J.E., Tielens, A.G.G.M., Whittet, D.C.B., Schutte, W.A., Boogert, A.C.A., Lutz, D., van Dishoeck, E.F., Bernstein, M.P. 2000, ApJ 537, 749 }
%
%
\rf{Chiar, J. E., Adamson, A. J., Pendleton, Y. J., Whittet, D. C. B., 
    Caldwell, D. A., Gibb, E. L., 2002, ApJ 570, 198 }
 
\rf{Cl\'enet, Y., Rouan, D., Gendron, E., Montri, J.,
 Rigaut, F., L\'na, P., Lacombe, F. 2001, A\&A, 376, 124 }
 
%
%
%
%
\rf{Cotera, A., Morris, M., Ghez, A. M., Becklin, E. E., Tanner, A. M., Werner, M. W., Stolovy, S. R. 1999, cpg conf, 240 }
\rf{de Graauw, T., Whittet, D.C.B. et al. 1996, A\&A 315,L345}
%
%
\rf{Duley, W. W., Williams, D. A. 1984, Natur 311, 685 }
 
\rf{Eckart, A. Moultaka, J. Viehmann, T. Straubmeier, C. Mouawad, N. 2004,
     ApJ 602, 760.}
 
%
%
 
\rf{Eckart, A, Ott, T, Genzel, R. 1999, A\&A 352,L22 }
 
%
%
%
%
%
%
%
 
\rf{Genzel, R., Thatte, N., Krabbe, A., Kroker, H., Tacconi-Garman, L. E. 1996, ApJ 472, 153 }
 
%
\rf{Genzel, R., Pichon, C., Eckart, A., Gerhard, O. \& Ott, T. 2000,
 Mon.Not.R.Soc.317, 348-374 }
\rf{Genzel, R., Sch\"odel, R., Ott, T., et al. 2003, ApJ. 594, 812-832 }
 
%
\rf{Gerakines, P. A., Whittet, D. C. B., Ehrenfreund, P., Boogert, A. C. A., Tielens, A. G. G. M., Schutte, W. A., Chiar, J. E., van Dishoeck, E. F., Prusti, T., Helmich, F. P., de Graauw, Th. 1999, ApJ 522, 357 }
%
%
%
%
%
%
%
%
%
 
\rf{Ghez, A., Duch\^ene, G., Matthews, K., et al. 2003, ApJ. 586,
   L127-L131 }
 
%
%
\rf{Grishko, V. I., Duley, W. W. 2002, ApJ 568, L131 }
%
%
%
%
%
\rf{Horrobin, M., Eisenhauer, F. et al. 2004, Astron. Nachr. 325, No. 2, 88-91 }

%
%
\rf{Hummer, D. G. \& Storey, P. J. 1987, MNRAS, 224, 801}
\rf{Ishii, M., Nagata, T., Sato, S.,
   Watanabe, M., Y., Yongqiang, J., Terry J. 1998, AJ 116, 868 }
%
%
\rf{Jones, T. J., Hyland, A. R., Allen, D. A. 1983, MNRAS 205, 187 }
%
%
%
%
%
 
\rf{Krabbe, A. et al. 1995, ApJL 447, L95 }
 

\rf{Lacy, J. H. \& Achtermann, J. M. 1991, ApJ 380, L71-L74}

\rf{Lebofsky, M. J. 1979, AJ 84, 324 }
 
%
\rf{Levin, Y. \& Beloborodov A. M. 2003, ApJ, 590, L33}
\rf{Lutz, D. et al. 1996, A\&A 315, 269 }
\rf{Maillard, J. P., Paumard, T., Stolovy, S. R., Rigaut, F. 2004, A\&A, 423, 155} 
%
%
%
\rf{McFadzean, A.D., Whittet, D.C.B., Bode, M.F., Adamson, A.J., Longmore, A.J. 1989, MNRAS 241, 873 }
\rf{Mennella, V., Baratta, G. A., Esposito, A., Ferini, G., 
   Pendleton, Y. J. 2003, ApJ 587, 727}
\rf{Morris, P.W., van der Hucht, K.A., Crowther, P.A., Hillier, D.J., Dessart, L., Williams, P.M., Willis, A.J., A\&A 353, 624}

%

\rf{Moultaka, J., Eckart, A., Viehmann, T., Straubmeier, C. Mouawad, N., Ott, T., Sch\"odel, R., 2004, A\&A 425, 529-542}
%
%
\rf{Najarro, F., Krabbe, A., Genzel, R., Lutz, D., Kudritzki, R. P., Hillier, D. J. 1997, A\&A 325, 700 }
%
%
\rf{Ott, T. 2004, PhD Thesis, Ludvig-Maximilians-Universit\"at, M\"unchen}
\rf{Paumard, T., Maillard, J.P., Morris, M., 
    Rigaut, F.  2001, A\&A 366, 466-480}

 \rf{Paumard, T., Genzel, R., Maillard, J.P., Ott, T., Morris, M., Eisenhauer, F., Abuter R. 2004, astro-ph/0407189}

 \rf{Paumard, T., Maillard, J.-P., Morris, M. 2004, A\&A 426, 81} 

\rf{Pendleton, Y.J., Sandford, S.A., Allamandola, L.J., Tielens, A.G.G.M., Sellgren, K. 1994, ApJ 437, 683 }
%
%
%
%
%
%
\rf{Rieke, G. H., Rieke, M. J., Paul, A. E. 1989, ApJ 336, 752 }
\rf{Rigaut, F., Geballe, T. R., Roy, J.-R., Draine, B. T. 2003, Astron. Nachr. 324, p. 551-555}
%
%
%
\rf{Sandford, S.A., Allamandola, L.J., Tielens, A.G.G.M., Sellgren, K., Tapia, M., Pendleton, Y. 1991, ApJ 371, 607 }
\rf{Sandford, S. A., Pendleton, Y. J., Allamandola, L. J. 1995, ApJ 440, 697-705}
\rf{Scoville, N.Z., Stolovy, S.R., Rieke, M., Christopher, M.H., 
   Yusef-Zadeh F. 2003 ApJ, 594, 294 }
 
\rf{Sch\"odel et al. 2002, Nature 419, 694-696 }
 
%
%
\rf{Sellgren, K., Brooke, T. Y., Smith, R. G., Geballe, T. R. 1995, ApJL 449, L69 }
%
%
%
%
%
%
%
\rf{Tanner, A., Ghez, A. M., Morris, M., Becklin, E. E., Cotera, A., 
    Ressler, M., Werner, M., Wizinowich, P. 2002, ApJ 575, 860}
 
\rf{Tanner, A., Ghez, A. M., Morris, M., Becklin, E. E. 2003, Astron. Nachr. 324, No. S1, 3-9}

\rf{Tanner, A., Ghez, A. M., Morris, M., Christou, J. C. 2004, astro-ph/0412494}
 
%
%
%
%
%
%
%
\rf{Viehmann, T., Eckart, A., Schoedel, R., Moultaka, J., Straubmeier, C., Pott, J.U. 2005, A\&A, in press}
\rf{Wada, S., Sakata, A., Tokunaga, A.T. 1991, ApJ 375, L17 }
%
%
%
%

\end{document}